\documentclass{PoS}

\def\refjl#1#2#3#4#5#6{\bibitem{#1} #2, { #3} {\bf #4} (#5) #6.}
\def\Journal#1#2#3#4{{#1} {\bf #2} (#4) #3}

\def\PRL{\em Phys. Rev. Lett.}

\def\NP{\em Nucl. Phys.}

\def\PL{\em Phys. Lett.}
\def\PR{\em Phys. Rev.}

\def\ARNPS{\em Ann. Rev. Nucl. Part. Sci.}

\newcommand{\eqn}[1]{(\ref{#1})}
\newcommand{\be}{\begin{equation}}
\newcommand{\ee}{\end{equation}}
\newcommand{\no}{\nonumber}
\newcommand{\bel}[1]{\be\label{#1}}
\newcommand{\ba}{\begin{array}{c}}
\newcommand{\bat}{\begin{array}{cc}}
\newcommand{\ea}{\end{array}}
\newcommand{\beqn}{\begin{eqnarray}}
\newcommand{\eeqn}{\end{eqnarray}}

\newcommand{\chpt}{$\chi$PT}
\newcommand{\rcht}{R$\chi$T}
\newcommand{\Frac}[2]{\frac{\displaystyle #1}{\displaystyle #2}}

\def\gap{\;\lower3pt\hbox{$\buildrel > \over \sim$}\;}
\def\lap{\;\lower3pt\hbox{$\buildrel < \over \sim$}\;}
\newcommand{\cL}{{\cal L}}

\newcommand{\cM}{{\cal M}}
\newcommand{\cO}{{O}}

\title{Low-Energy Constants from Resonance Chiral Theory }

\ShortTitle{Low-Energy Constants from \rcht}

\author{Antonio Pich\\ 
        Departament de F\'{\i}sica Te\`orica, IFIC,
Universitat de Val\`encia--CSIC,\\
Apt. Correus 22085, E-46071 Val\`encia, Spain\\
        E-mail: \email{Antonio.Pich@ific.uv.es}}

\abstract{I discuss the recent attempts to build an
effective chiral Lagrangian incorporating massive resonance states.
A useful approximation scheme to organize the resonance Lagrangian is provided
by the large-$N_C$ limit of QCD.
Integrating out the resonance fields, one recovers the usual chiral perturbation theory
Lagrangian with explicit values for the low-energy constants, parameterized in terms of resonance
masses and couplings.
The resonance chiral theory generates Green functions that interpolate between
QCD and chiral perturbation theory. Analyzing these Green functions, both for large and small
momenta, one gets QCD constraints on the resonance couplings and, therefore, information on the
low-energy constants governing the Goldstone interactions.}

\FullConference{8th Conference Quark Confinement and the Hadron Spectrum \\
		 September 1-6 2008\\
		 Mainz, Germany}

\begin{document}

\section{Chiral Symmetry}
\label{sec:chpt}

With $n_f$ massless quark flavours, the QCD Lagrangian
is invariant under global \ $SU(n_f)_L\otimes SU(n_f)_R$
transformations of the left- and right-handed quarks in flavour space.
The symmetry group spontaneously breaks down to the diagonal
subgroup $SU(n_f)_{L+R}$ \ and \
$n_f^2-1$ pseudoscalar massless Goldstone bosons appear in the theory,
which for $n_f=3$ can be identified with the
eight lightest hadronic states $\phi^a =\{\pi$, $K$, $\eta\}$.
These pseudoscalar fields are usually parameterized through the
$3\times 3$ unitary matrix \
$U(\phi) =  u(\phi)^2 =\exp{\left\{i\lambda^a\phi^a/f\right\}}$.

The Goldstone nature of the pseudoscalar mesons implies strong
constraints on their interactions, which can be most easily analyzed
on the basis of an effective Lagrangian containing only the Goldstone modes
\cite{WE:79,GL:85,ChPTrevs}.
The low-energy effective Lagrangian is organized in
terms of increasing powers of momenta (derivatives) and quark masses: \
$\cL = \sum_n \cL_{2n}$.
At lowest order, the most general effective Lagrangian
consistent with chiral symmetry has the form \cite{GL:85}:
\bel{eq:lowestorder}
\cL_2 = {f^2\over 4}\,
\langle D_\mu U^\dagger D^\mu U \, + \, U^\dagger\chi  \,
+  \,\chi^\dagger U\rangle\, ,
\qquad\qquad
\chi \equiv 2 B_0 \, (s + i p) \, ,
\ee
where \
$D_\mu U = \partial_\mu U -i r_\mu U + i U\, l_\mu$ \  and
$\langle A\rangle$ denotes the flavour trace of the matrix $A$.
The external Hermitian matrix-valued sources
$l_\mu$, $r_\mu$, $s$ and $p$\ are used to generate the corresponding
left, right, scalar and pseudoscalar
QCD Green functions and allow to incorporate the
explicit breaking of chiral symmetry through the quark masses:
$s = \cM + \ldots\, ,\,\cM= \hbox{\rm diag}(m_u,m_d,m_s)$.
The constants $B_0$ and $f$ are not fixed by symmetry requirements;
one finds that $f$ equals the pion decay constant
(at lowest order) $f = f_\pi = 92.3$~MeV, while $B_0$ is
related to the quark condensate:
\bel{eq:B0}
B_0 = -{\langle\bar q q \rangle\over f^2} =
{M_\pi^2\over m_u + m_d} = {M_{K^0}^2\over m_s + m_d} =
{M_{K^\pm}^2\over m_s + m_u}\, .
\ee
With only two low-energy constants, the lowest-order chiral
Lagrangian $\cL_2$ encodes in a very compact way all the
Current Algebra results obtained in the sixties.

The symmetry constraints become less powerful at higher orders.
At $O(p^4)$ we need ten additional coupling constants $L_i$
to determine the low-energy behaviour of the Green functions \cite{GL:85}:
\bel{eq:l4}
\cL_4  = L_1 \,\langle D_\mu U^\dagger D^\mu U\rangle^2 \, + \,
L_2 \,\langle D_\mu U^\dagger D_\nu U\rangle\,
  \langle D^\mu U^\dagger D^\nu U\rangle\, + \,\ldots
\ee
One-loop graphs with the lowest-order Lagrangian $\cL_2$ contribute
also at $O(p^4)$.
Their divergent parts are renormalized by the $\cL_4$ couplings,
which introduces a renormalization-scale dependence.
The chiral loops generate non-polynomial contributions,
with logarithms and threshold factors as required by unitarity,
which are completely determined as functions of $f$ and the Goldstone
masses.

The precision required in present phenomenological applications
makes necessary to include corrections of $O(p^6)$ \cite{BI:07}.
This involves contributions from $\cL_4$ at one-loop and
$\cL_2$ at two-loops, which can be fully predicted \cite{bce00}.
However, the $O(p^6)$ chiral Lagrangian $\cL_6$ contains
90 (23) independent local terms of even (odd)
intrinsic parity \cite{bce00,BGP:01,Fearing:1994ga}.
The huge number of unknown couplings limits the achievable accuracy.
Clearly, further progress will depend on our ability
to estimate these chiral couplings, which encode the underlying QCD dynamics.


\section{Resonance Chiral Theory}
\label{sec:RChT}

The limit of an infinite number of quark colours
is a very useful starting point to understand many
features of QCD \cite{HO:74,WI:79}.
Assuming confinement,
the strong dynamics at $N_C\to\infty$ is given
by tree diagrams with infinite sums of hadron exchanges,
which correspond to the tree approximation to some local
effective Lagrangian. Hadronic loops generate corrections
suppressed by factors of $1/N_C$.
At $N_C\to\infty$, QCD has a larger symmetry
$U(3)_L\otimes U(3)_R\to U(3)_{L+R}$, and
one needs to include in the matrix $U(\phi)$ a ninth
Goldstone boson field, the $\eta_1$.
Resonance chiral theory (\rcht) \cite{EGPR:89,EGLPR:89,RChTc} provides
an appropriate framework to incorporate the massive mesonic states \cite{PI:02}.

Let us consider a chiral-invariant Lagrangian describing the couplings of resonance nonet multiplets
$V_i^{\mu\nu}(1^{--})$, $A_i^{\mu\nu}(1^{++})$, $S_i(0^{++})$ and $P_i(0^{-+})$ to
the Goldstone bosons.
At lowest order in derivatives the interaction Lagrangian $\cL_R$, linear in the resonance fields,
takes the form \cite{EGPR:89}:
\beqn\label{eq:L_R}
\cL_R &=& \sum_i\;\biggl\{ {F_{V_i}\over 2\sqrt{2}}\;
\langle V_i^{\mu\nu} f_{+ \, \mu\nu}\rangle\, +\,
{i\, G_{V_i}\over \sqrt{2}} \,\,\langle V_i^{\mu\nu} u_\mu u_\nu\rangle
\, +\,
{F_{A_i}\over 2\sqrt{2}} \;\langle A_i^{\mu\nu} f_{- \, \mu\nu} \rangle
\biggr.\no\\ &&\hskip .5cm\biggl.\mbox{} +\,
c_{d_i} \; \langle S_i\, u^\mu u_\mu\rangle
\, +\, c_{m_i} \; \langle S_i\, \chi_+ \rangle
\, +\, i\, d_{m_i}\;\langle P_i\, \chi_- \rangle
\biggr\}\, ,
\eeqn
where \
$u_\mu \equiv i\, u^\dagger D_\mu U u^\dagger$, \
$f^{\mu\nu}_\pm\equiv u F_L^{\mu\nu} u^\dagger\pm  u^\dagger F_R^{\mu\nu} u$
\ with $F^{\mu\nu}_{L,R}$ the field-strength tensors of the
$l^\mu$ and $r^\mu$ flavour fields
\ and \
$\chi_\pm\equiv u^\dagger\chi u^\dagger\pm u\chi^\dagger u$.
The resonance couplings
$F_{V_i}$, $G_{V_i}$, $F_{A_i}$, $c_{d_i}$, $c_{m_i}$ and $d_{m_i}$
are of \ $O\left(\sqrt{N_C}\,\right)$.

The lightest resonances have an important impact on the
low-energy dynamics of the pseudoscalar bosons.
Below the resonance mass scale, the singularity associated with the
pole of a resonance propagator is replaced by the corresponding
momentum expansion; therefore, the exchange of virtual resonances generates
derivative Goldstone couplings proportional to powers of $1/M_R^2$.
At lowest order in derivatives, this gives the large--$N_C$ predictions
for the $O(p^4)$ couplings of chiral perturbation theory (\chpt) \cite{EGPR:89}:
%
$$
2\, L_1 = L_2 = \sum_i\; {G_{V_i}^2\over 4\, M_{V_i}^2}\, ,  \qquad
L_3 = \sum_i\;\left\{ -{3\, G_{V_i}^2\over 4\, M_{V_i}^2} +
{c_{d_i}^2\over 2\, M_{S_i}^2}\right\} \, ,
\qquad
L_5 = \sum_i\; {c_{d_i}\, c_{m_i}\over M_{S_i}^2} \, ,
$$
\be\label{eq:vmd_results}
L_8 = \sum_i\;\left\{ {c_{m_i}^2\over 2\, M_{S_i}^2} -
{d_{m_i}^2\over 2\, M_{P_i}^2}\right\} \, , \qquad
L_9 = \sum_i\; {F_{V_i}\, G_{V_i}\over 2\, M_{V_i}^2}\, ,
\qquad
L_{10} = \sum_i\;\left\{ {F_{A_i}^2\over 4\, M_{A_i}^2}
 - {F_{V_i}^2\over 4\, M_{V_i}^2}\right\}  \, .
\ee
All these couplings are of $O(N_C)$, in agreement with the
counting indicated in Table~\ref{tab:Lcouplings}, while for the
couplings of $O(1)$ we get: \
$2\, L_1-L_2 = L_4 = L_6 = L_7 = 0$.

\begin{table}[tbh]\centering
\renewcommand{\arraystretch}{1.1}
\begin{tabular}{|c|c|c|c|c|}
\hline
$i$ & \raisebox{0pt}[10pt][5pt]{$L_i^r(M_\rho)$} & $O(N_C^m)$  & Source &
$L_{i}^{N_C\to\infty}$
\\
\hline
\raisebox{0pt}[10pt]{$2L_1-L_2$}
& $-0.6\pm0.6$ & $O(1)$ &
$K_{e4}$, $\pi\pi\to\pi\pi$ & 0
\\
$L_2$ & $\hphantom{-}1.4\pm0.3$ & $O(N_C)$ &
$K_{e4}$, $\pi\pi\to\pi\pi$ & $\phantom{-}1.8$
\\
$L_3$ & $-3.5\pm1.1$ & $O(N_C)$ & $K_{e4}$, $\pi\pi\to\pi\pi$ & $-4.3$
\\
$L_4$ & $-0.3\pm0.5$ & $O(1)$ & Zweig rule & 0
\\
$L_5$ & $\hphantom{-}1.4\pm0.5$ & $O(N_C)$ & $F_K : F_\pi$ & $\phantom{-}2.1$
\\
$L_6$ & $-0.2\pm0.3$ & $O(1)$ & Zweig rule & 0
\\
$L_7$ & $-0.4\pm0.2$ & $O(1)$ & GMO, $L_5$, $L_8$ & $-0.3$
\\
$L_8$ & $\hphantom{-}0.9\pm0.3$ & $O(N_C)$ & $M_\phi$, $L_5$ & $\phantom{-}0.8$
\\
$L_9$ & $\hphantom{-}6.9\pm0.7$ & $O(N_C)$ &
$\langle r^2\rangle^\pi_V$ & $\phantom{-}7.1$
\\
$L_{10}$ & $-5.5\pm0.7$ & $O(N_C)$ & $\pi\to e\nu\gamma$ & $-5.4$
\\[1pt]\hline
\end{tabular}
\caption{Phenomenological values [$O(p^4)$] of the renormalized couplings
$L_i^r(M_\rho)$ in units of $10^{-3}$.
The large--$N_C$ predictions obtained within the single-resonance
approximation are given in the last column.} 
\label{tab:Lcouplings}
\end{table}

Owing to the $U(1)_A$ anomaly, the $\eta_1$ field is massive and it is often
integrated out from the low-energy chiral theory. In that case,
the $SU(3)_L\otimes SU(3)_R$ chiral coupling $L_7$ gets a contribution
from $\eta_1$ exchange \cite{GL:85,EGPR:89}:
\bel{eq:L7}
L_7 = - {f^2\over 48\, M^2_{\eta_1}} \, .
\ee

\section{Short-Distance Constraints}

The short-distance properties of the underlying QCD dynamics
impose some constraints on the resonance
parameters \cite{EGLPR:89,PI:02}.
At leading order in $1/N_C$, the two-Goldstone matrix element of the
vector current is characterized by the vector form factor
\bel{eq:VFF}
F_V(t)\, =\, 1\, + \, \sum_i\,
{F_{V_i}\, G_{V_i}\over f^2}\; {t\over M_{V_i}^2-t} \, .
\ee
Since  $F_V(t)$ should vanish at infinite momentum
transfer $t$, the resonance couplings should satisfy
\bel{eq:SD1}
\sum_i\, F_{V_i}\, G_{V_i}\, =\, f^2\, .
\ee
Similarly,
the matrix element of the axial current between one Goldstone and
one photon is parameterized by the so-called axial form factor $G_A(t)$,
which vanishes at $t\to\infty$ provided that
\bel{eq:SD2}
\sum_i\,
\left(2\, F_{V_i}\, G_{V_i}- F_{V_i}^2\right) / M_{V_i}^2
\, =\, 0\, .
\ee
Requiring the scalar form factor $F^S(t)$, which governs the
two-pseudoscalar matrix element of the scalar quark current,
to vanish at $t\to\infty$, one gets the constraints \cite{JOP:02}:
\bel{eq:SD4}
4\,\sum_i\,c_{d_i}\, c_{m_i} = f^2 \, ,
\qquad\qquad
\sum_i\,  c_{m_i}\,\left( c_{m_i}-c_{d_i}\right) / M_{S_i}^2 = 0 \, .
\ee

Since gluonic interactions preserve chirality,
the two-point function built from a left-handed and a right-handed
vector quark currents
$\Pi_{LR}(t)$ satisfies an unsubtracted dispersion relation.
In the chiral limit, it vanishes faster than $1/t^2$
when $t\to\infty$; this implies the well-known Weinberg conditions \cite{WE:67}:
\bel{eq:SD3}
\sum_i\,\left( F_{V_i}^2 - F_{A_i}^2\right) = f^2 \, ,
\qquad\qquad
\sum_i\,\left( M_{V_i}^2 F_{V_i}^2 - M_{A_i}^2 F_{A_i}^2\right)
= 0 \, .
\ee

The two-point correlators of two scalar or two pseudoscalar
currents would be equal if chirality was preserved.
For massless quarks, $\Pi_{SS-PP}(t)$ vanishes as $1/t^2$ when
$t\to\infty$, with a coefficient proportional to  
$\alpha_s\,\langle\bar q\Gamma q\,\bar q\Gamma q\rangle
\sim\alpha_s\,\langle\bar q q\rangle^2 \sim \alpha_s\, B_0^2$.
Imposing this behaviour, one gets \cite{GP:00}:
\bel{eq:SD5}
8\,\sum_i\left( c_{m_i}^2 - d_{m_i}^2\right) = f^2  ,
\qquad\qquad
\sum_i\left( c_{m_i}^2 M_{S_i}^2 - d_{m_i}^2 M_{P_i}^2\right) =
3\,\pi\alpha_s\, f^4 /4\, .
\ee

\section{Single-Resonance Approximation}

Let us approximate each infinite resonance sum with
the first meson-nonet contribution.
This is meaningful at
low energies where the contributions from higher-mass states are
suppressed by their corresponding propagators.
The resulting short-distance constraints are
matching conditions between an effective theory
below the scale of the second resonance multiplets
and the underlying QCD dynamics.
With this approximation, Eqs.~\eqn{eq:SD1}, \eqn{eq:SD2} and \eqn{eq:SD3}
determine the vector and axial-vector couplings in terms of $M_V$
and $f$ \cite{EGLPR:89}:
\bel{eq:VA_coup}
F_V = 2\, G_V = \sqrt{2}\, F_A = \sqrt{2}\, f \, ,
\qquad\qquad
M_A = \sqrt{2}\, M_V \, .
\ee
The scalar \cite{JOP:02}
and pseudoscalar parameters are obtained
from \eqn{eq:SD4} and \eqn{eq:SD5} \cite{PI:02}:
\bel{eq:SP_coup}
c_m = c_d = \sqrt{2}\, d_m = f/2 \, ,
\qquad\qquad
M_P = \sqrt{2}\, M_S \, \left(1 - \delta\right)^{1/2}\, .
\ee
The last relation involves a small correction \
$\delta \approx 3\,\pi\alpha_s f^2/M_S^2 \sim 0.08\,\alpha_s$,
which we can neglect together with the tiny effects from
light quark masses.

Inserting these predictions into Eqs.~\eqn{eq:vmd_results},
one finally gets all $O(p^4 N_C)$ \chpt\
couplings, in terms of $M_V$, $M_S$ and $f$:
\bel{eq:Li_SRA_1}
2\, L_1 = L_2 = \frac{1}{4}\, L_9 = -\frac{1}{3}\, L_{10}
= {f^2\over 8\, M_V^2}\, ,
\ee
\bel{eq:Li_SRA_2}
L_3 = -{3\, f^2\over 8\, M_V^2} + {f^2\over 8\, M_S^2}\, ,
\qquad\quad
L_5 ={f^2\over 4\, M_S^2}\, ,
\qquad\quad
L_8 = {3\, f^2\over 32\, M_S^2}\, .
\ee
The last column in Table~\ref{tab:Lcouplings} shows the
results obtained with $M_V = 0.77$~GeV,
$M_S = 1.0$~GeV and $f=92$~MeV. Also shown is the $L_7$
prediction in \eqn{eq:L7}, taking
$M_{\eta_1} = 0.80$~GeV. The agreement with
the measured values is a clear
success of the large--$N_C$ approximation.
It demonstrates that the lightest resonance multiplets
give indeed the dominant contributions at low energies.

Corrections induced by \rcht\ couplings quadratic in the resonance fields have been considered \cite{RSP:07,RSP:08}.
Although they slightly modify some of the previous relations, the general pattern remains so that all $O(p^4 N_C)$
\chpt\ couplings are still successfully determined in terms of resonance masses and the pion decay constant.
The possible effect of more exotic $2^{++}$ and $1^{+-}$ resonance exchanges has been analyzed recently.
The short-distance constraints eliminate any possible contribution to the $L_i$ couplings
from $1^{+-}$ exchange and only allow a tiny $2^{++}$ contribution to $L_3$,  $L_3^T=0.16\cdot 10^{-3}$, which is
negligible compared to the sum of vector and scalar contributions \cite{EZ:07}.
This small tensor contribution had been previously obtained in the SU(2) theory \cite{SU2-T}.

The study of other Green functions provides further matching
conditions between the hadronic and fundamental QCD descriptions.
Clearly, it is not possible to satisfy all of them
within the single-resonance approximation, since QCD requires an infinite number of massive states. A useful generalization is the
so-called {\it Minimal Hadronic Ansatz}, which
keeps the minimum number of resonances compatible with all known
short-distance constraints for the problem at hand \cite{KPdR}.

\section{Determination of $\mathbf{O(p^6)}$ Low-Energy Couplings}

The most general \rcht\ Lagrangian contributing to the $O(p^6)$ \chpt\ couplings
has been recently constructed in Ref.~\cite{RChTc}. A priori the Lagrangian contains
a long list of possible operators, including terms with one [$O(p^4)$], two [$O(p^2)$] and
three [$O(p^0)$] resonance fields. Many of them can be eliminated, using the equations of motion,
field redefinitions and algebraic identities. The functional integration of the resonance
fields has been completed, obtaining the large--$N_C$ resonance contributions to all $O(p^6)$ \chpt\ couplings $C_i$
in terms of resonance parameters. Those low-energy constants which don't get any resonance contribution
have been identified and useful relations among different couplings have been obtained.
However, there remain still many unknown resonance parameters which require a further investigation
of short-distance QCD constraints. A complete matching between QCD and \rcht\ has not yet been achieved at
this order.

Some $O(p^6)$ \chpt\ couplings have been already determined by studying an appropriate set of three-point
functions \cite{MO:95,KN:01,RPP:03,CEEPP:04,CEEKPP:05,BGLP:03}. For instance, the analysis of the $\langle VAP\rangle$
Green function allows to derive the values \cite{RChTc,CEEPP:04}:
\begin{equation}
 \label{LowEnergySolution}
\begin{tabular}{lll}
$C_{78} = \displaystyle\frac{f^2(3 M_A^2 + 4 M_V^2)}{8 M_V^4 M_A^2}
- \displaystyle\frac{f^2}{16 M_V^2 M_P^2}$  , & \hspace*{.1cm} &
$C_{82} = - \displaystyle\frac{f^2(4 M_A^2 + 5 M_V^2)}{32 M_V^4
  M_A^2}
- \displaystyle\frac{f^2}{32 M_A^2 M_P^2}$  ,  \\[.3cm]
$C_{87} = \displaystyle\frac{f^2(M_A^4 + M_V^4 + M_A^2 M_V^2)}
{8 M_V^4 M_A^4}$  ,  & \hspace*{.1cm} &
$C_{88} = - \displaystyle\frac{f^2}{4 M_V^4} +
\displaystyle\frac{f^2}{8 M_V^2 M_P^2}$  ,  \\[.3cm]
$C_{89} = \displaystyle\frac{f^2(3 M_A^2 + 2 M_V^2)}{4 M_V^4 M_A^2}$
 ,  & \hspace*{.1cm} &
$C_{90} = \displaystyle\frac{f^2}{8 M_V^2 M_P^2}$  .
\end{tabular}
\end{equation}
From a similar analysis of the $\langle SPP\rangle$ Green function, one obtains \cite{RChTc,CEEKPP:05}:

\be\label{Lecs-model-2}
C_{12} = - \frac{f^2}{8 M_S^4}\, ,
\qquad
C_{34} =  \frac{3 \, f^2}{16 M_S^4} + \frac{f^2}{16}
\left(\frac{1}{M_S^2} - \frac{1}{M_P^2} \right)^2\, ,
\qquad
C_{38}  = \frac{f^2}{8 M_S^4} - \frac{f^2}{16 M_P^4}\, .
\ee
The couplings $C_{12}$ and $C_{34}$ govern the amount of SU(3) breaking in the $K_{l3}$ form factor at zero momentum transfer
and, therefore, have important implications in the determination of $|V_{us}|$ \cite{CEEKPP:05}.

\section{Subleading $\mathbf{1/N_C}$ Corrections}

The large--$N_C$ limit provides a very successful description
of the low-energy dynamics \cite{PI:02}.
However, we are still lacking a systematic procedure to incorporate contributions of
next-to-leading order (NLO) in the $1/N_C$ counting.
The first efforts concentrated in pinning down the most relevant
subleading effects, such as the resonance widths which regulate the corresponding poles
in the meson propagators \cite{GP:97}, or the role of final state interactions in
the physical amplitudes \cite{JOP:02,GP:97,PaP:01,IAM}.

More recently, methods to determine the low-energy constants of \chpt\ at the next-to-leading order
in $1/N_C$ have been developed \cite{RSP:07,RSP:08,CP:01,RSP:04}. This is an important issue because
the dependence of the \chpt\ couplings with the renormalization scale is a subleading effect in the $1/N_C$ counting.
Since the usual resonance-saturation estimates have been performed at $N_C\to\infty$, they are unable to control
the renormalization-scale dependence of the low-energy couplings (at which value of $\mu$ the estimates apply?).

Quantum loops including virtual resonance propagators constitute a major technical challenge.
Their ultraviolet divergences require higher-dimensional
counterterms, which could generate a problematic behaviour at large momenta \cite{RSP:04}.
Thus, it is necessary to investigate
the short-distance QCD constraints at the next-to-leading order in $1/N_C$.
A first step in this direction was achieved through a one-loop calculation of the vector
form factor in the \rcht\ \cite{RSP:04}, which demonstrated that the matching
with the underlying QCD dynamics strongly constrains the ultraviolet behaviour of \rcht, determining
the renormalized couplings needed for this particular calculation. This fact appears to be quite general \cite{SC:07}
and has been further corroborated through a recent investigation
of the full one-loop generating functional that arises from
\rcht\ with only one multiplet of scalar and pseudoscalar resonances \cite{PRR:07}.

Using analyticity and unitarity, it is possible to avoid all technicalities associated with the renormalization
procedure, reducing the calculation of one-loop Green functions to tree-level diagrams plus dispersion
relations \cite{RSP:07,RSP:08}. This allows
to understand the underlying physics in a much more transparent way. In particular, the subtle
cancellations among many unknown renormalized couplings found in \cite{RSP:04} and the relative
simplicity of the final result can be better understood in terms of the imposed short-distance
constraints.

As an example, let us consider the difference between the vector and axial-vector two-point
functions $\Pi_{V-A}(t)\equiv \Pi_{VV}(t)-\Pi_{AA}(t)$. Its low-energy behaviour is dictated by \chpt\ 
\cite{GL:85,bce00,op6-correlator}:
\be\label{eq:Pi_chpt}
  \Pi_{V-A}(t) =  \frac{2 f^2}{t}  -  8 L_{10}^r(\mu)
 - \frac{\Gamma_{10}}{4\pi^2} \left( \Frac{5}{3}-\ln \frac{-t}{\mu^2} \right) +
t \left[ 16\,C_{87}^r (\mu)  -\frac{\Gamma_{87}^{(L)}}{2\pi^2f^2}
\left( \Frac{5}{3}-\ln \frac{-t}{\mu^2} \right) \right] + \cO\left(N_C^{0}t\right) \, ,
\ee
with $\Gamma_{10} = -1/4$ and  $\Gamma_{87}^{(L)} = - L_9^r(\mu)/2$. The couplings $f^2$, $L_{10}$ and $C_{87}$ are of $\cO(N_C)$, while $\Gamma_{10}$ and $\Gamma_{87}^{(L)}/f^2$ are of $\cO(N_C^0)$ and represent a NLO effect.
The term $2f^2/t$ contains the pole generated by the Goldstone-boson exchange.
In the large--$N_C$ limit, $\Pi_{V-A}(t)$ receives in addition tree-level contributions from vector and axial-vector exchanges, which are easily
computed within \rcht. Expanding the \rcht\ expression in powers of momenta, one recovers the resonance-exchange predictions for the low-energy
couplings $L_{10}$ and $C_{87}$ in Eqs.~\eqn{eq:Li_SRA_1} and \eqn{LowEnergySolution}.

At NLO in $1/N_C$, $\Pi_{V-A}(t)$ contains one-loop contributions from two-body exchanges of Goldstone bosons and heavy resonances, which give rise to ultraviolet divergences. However, these loop corrections can be fully determined from their finite absorptive contributions, through dispersive relations.
The ultraviolet behaviour is then parameterized through the corresponding subtraction constants, which are fixed by the short-distance
QCD behaviour requiring the correlator to vanish faster than $1/t^2$ at infinite momentum. The contributions from the dominant $\pi\pi$, $\pi V$,
$\pi A$, $\pi S$ and $\pi P$ exchanges have been computed in Ref.~\cite{RSP:08}. It is remarkable that, imposing a good short-distance behaviour for the corresponding vector and axial-vector spectral functions, one fully determines the relevant contributing form factors within the single resonance approximation.
The low momentum expansion of the resulting $\Pi_{V-A}(t)$ correlator reproduces Eq.~\eqn{eq:Pi_chpt}, with explicit values for
$L_{10}^r(\mu)$ and $C_{87}(\mu)$ which only depend on the resonance masses and the pion decay constant. The logarithmic dependence with the \chpt\
renormalization scale is fully reproduced through the Goldstone loops.
%
\begin{figure}[tbh]
\begin{center}
\begin{minipage}[c]{0.47\linewidth}\centering
\includegraphics[clip,scale=0.67]{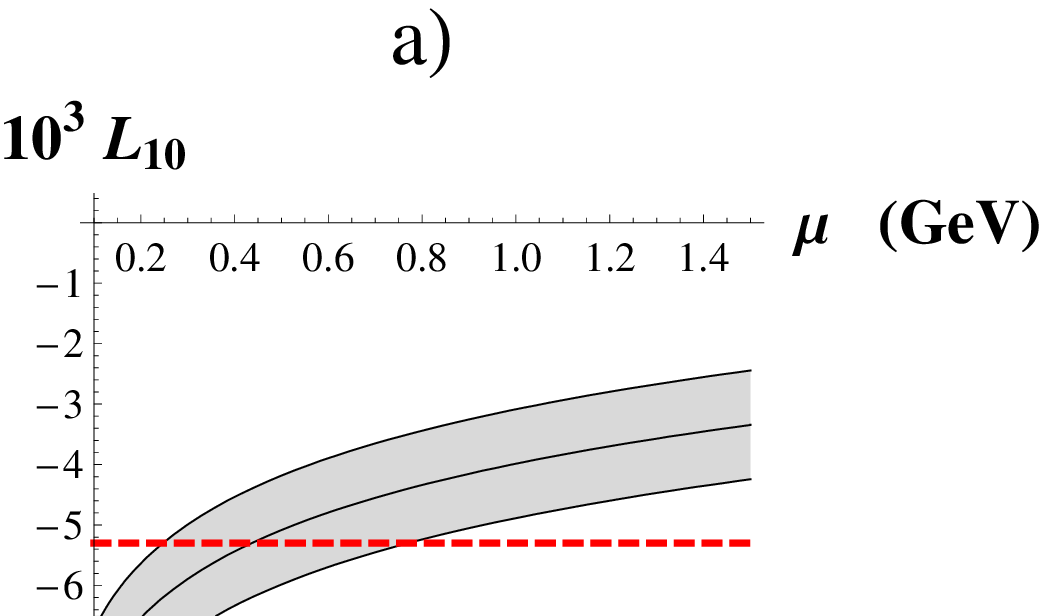}
\end{minipage}
\hfill
\begin{minipage}[c]{0.47\linewidth}\centering
\vskip .3cm
\includegraphics[clip,scale=0.67]{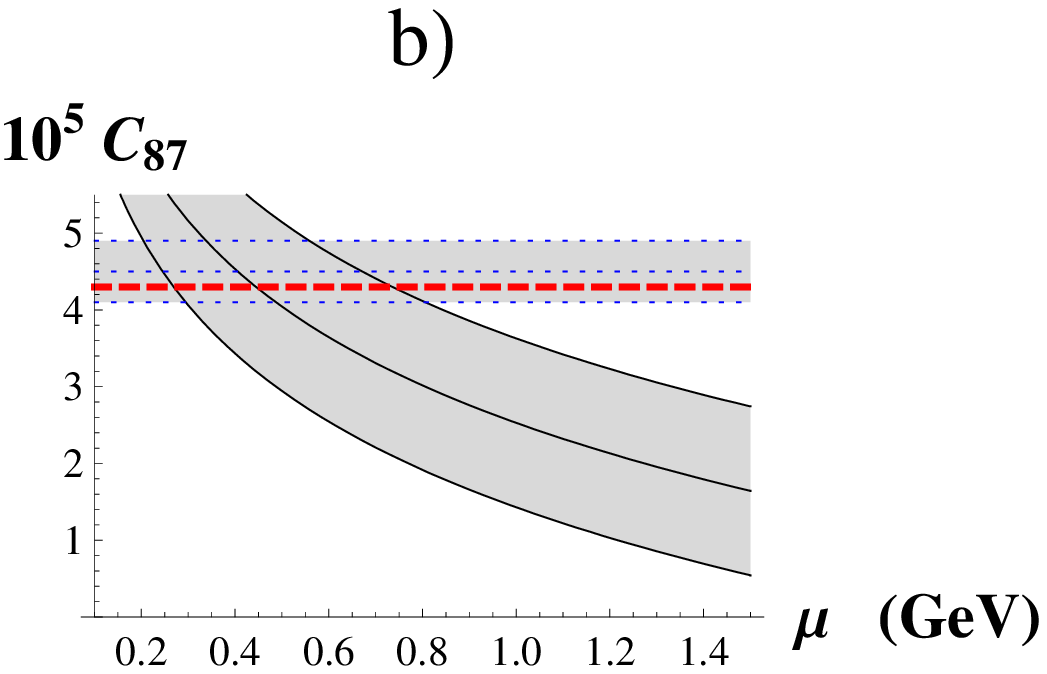}
\end{minipage}\vskip .2cm
\caption{\label{run-LECs}
NLO predictions (solid gray bands) for $L_{10}^r(\mu)$ (left) and $C_{87}^r(\mu)$ (right, $1/f^2$ units),
compared to the LO estimates (dashed) and the result from Ref.~\cite{C87-Peris} (dotted) obtained with Pade approximants.
}
\end{center}
\end{figure}
%
The resulting predictions for the two low-energy constants as functions of the \chpt\ renormalization scale are shown in Fig.~\ref{run-LECs}.
At the reference scale $\mu_0=770~\mathrm{MeV}$, one gets the numerical values \cite{RSP:08}:
\be
 L_{10}^r(\mu_0) = \left(-4.4 \pm 0.9\right)\cdot 10^{-3} \, ,
\qquad\qquad
C_{87}^r(\mu_0) = \left(3.9 \pm 1.4\right)\cdot 10^{-3}\:\mathrm{GeV}^{-2}\, ,
\ee
where the uncertainties reflect the present errors associated with the input resonance masses.
These numbers are in very good agreement with the recent and more precise $\cO(p^6)$ phenomenological
determination of these constants from $\tau$-decay data:
$L_{10}^r(\mu_0) = \left(-4.06 \pm 0.39\right)\cdot 10^{-3}$
and $C_{87}^r(\mu_0) = \left(4.89 \pm 0.19\right)\cdot 10^{-3}\:\mathrm{GeV}^{-2}$
\cite{GPP:08}.

The difference between the scalar and pseudoscalar two-point functions, $\Pi_{S-P}(t)\equiv \Pi_{SS}(t)-\Pi_{PP}(t)$,
has been also analyzed within \rcht, at the NLO, in a completely analogous way \cite{RSP:07}. Once more,
the short-distance QCD constraints are able to fix all relevant resonance couplings in terms of the pion decay constant and
resonance masses. The corresponding low-energy expansion of $\Pi_{S-P}(t)$ provides then a determination of the \chpt\
couplings $L_{8}^r(\mu)$ and $C_{38}^r(\mu)$ at the NLO in $1/N_C$, keeping full control of the renormalization-scale dependence.
At the reference scale $\mu_0$, one gets the values \cite{RSP:07}:
\be
L_{8}^r(\mu_0) = \left(0.6 \pm 0.4\right)\cdot 10^{-3} \, ,
\qquad\qquad
C_{38}^r(\mu_0) = \left(0.3 \pm 0.8\right)\cdot 10^{-3}\:\mathrm{GeV}^{-2}\, .
\ee
The predicted value for $L_8$ is in good agreement with the $\cO(p^6)$ phenomenological determination
$L_{8}^r(\mu_0) = \left(0.62 \pm 0.20\right)\cdot 10^{-3}$ \cite{ABT:01}.

\section{Summary}

The $1/N_C$ expansion provides a useful bridge between short and long distances and a powerful power-counting parameter.
The strong dynamics at $N_C\to\infty$ corresponds to the tree approximation to some local effective Lagrangian (with an infinite number of degrees of freedom). \rcht\ constitutes an appropriate effective Lagrangian implementation of the large--$N_C$ world, incorporating the chiral symmetry constraints. It allows to obtain useful approximations to the QCD Green functions, in terms of a finite number of meson fields, which interpolate between \chpt\ and the underlying QCD theory.

Integrating out the heavy resonance fields one recovers at low energies
the \chpt\ Lagrangian with explicit values of the chiral couplings in terms of resonance
parameters. Since the short-distance properties of QCD impose stringent constraints
on the \rcht\ couplings, it is then possible to extract information on the low-energy
constants of \chpt.

Truncating the infinite tower of meson resonances to the lowest states with $0^{-+}$,
$0^{++}$, $1^{--}$ and $1^{++}$ quantum numbers one gets
a very successful prediction of the $\cO(p^4 N_C)$ \chpt\ couplings in terms of only three
parameters: $M_V$, $M_S$ and the pion decay constant $f$. This provides a theoretical
understanding of the role of resonance saturation in low-energy phenomenology, which has been recently extended to
$\cO(p^6)$.

Hadronic loops generate corrections suppressed by factors of $1/N_C$, which can be analyzed
within \rcht. The short-distance QCD constraints turn out to be crucial in order to
control the ultraviolet behaviour of the effective theory; together with analyticity and unitarity,
they allow to determine the Green functions at the NLO in $1/N_C$. Taking the low-energy limit, it is then possible to
pin down the \chpt\ couplings at NLO and, therefore, to control their chiral renormalization-scale dependence.
Only a few explicit calculations have been done up to now, with very successful results. Further progress is to be expected in the
near future \cite{progress}.


\section*{Acknowledgements}
I'm grateful to J. Portol\'es, I. Rosell and J.J. Sanz-Cillero for useful comments on the manuscript.
This work has been supported
by MICINN, Spain (grants FPA2007-60323 and
Consolider-Ingenio 2010 CSD2007-00042, CPAN), by the
EU Contract MRTN-CT-2006-035482 (FLAVIAnet) and by Generalitat Valenciana
(PROMETEO/2008/069).


\end{document}